\documentclass[11pt,twoside]{article}
\usepackage{asp2010}

\resetcounters

\begin{document}

\title{A Tale of Two Stars: Interferometric Studies of Post-AGB Binaries}
\author{M. Hillen$^1$, J. Menu$^1$, B. de Vries$^2$, H. Van Winckel$^1$, 
M. Min$^3$, G.D. Mulders$^4$, C. Gielen$^5$, T. Wevers$^6$, S. Regibo$^1$, 
and T. Verhoelst$^5$
\affil{$^1$Instituut voor Sterrenkunde, University of Leuven, Leuven, 
Belgium}
\affil{$^2$Department of Astronomy, AlbaNova University Center, Stockholm
University, Stockholm, Sweden}
\affil{$^3$Sterrenkundig Instituut Anton Pannekoek, University of Amsterdam, 
Amsterdam, The Netherlands}
\affil{$^4$Lunar and Planetary Laboratory, The University of Arizona, 
Tucson AZ, USA}
\affil{$^5$Belgian Institute for Space Aeronomy, Brussels, Belgium}
\affil{$^6$Department of Astrophysics\,/\,IMAPP, Radboud University 
Nijmegen, Nijmegen, The Netherlands}
}

\begin{abstract}
Binaries with circumbinary disks are commonly found among 
optically bright post-AGB stars.  Although clearly linked to binary 
interaction processes, the formation, evolution and fate of these disks 
are still badly understood.  Due to their compactness, interferometric 
techniques are required to resolve them. Here, we discuss our 
high-quality multiwavelength interferometric data of two prototypical 
yet very different post-AGB binaries, AC and 89 Herculis, as well 
as the modeling thereof with radiative transfer models. A detailed 
account of the data and models of both objects is published in three 
separate papers elsewhere; here we focus on comparing the modeling 
results for the two objects. In particular we discuss the successes 
and limitations of the models which were developed for protoplanetary 
disks around young stars. We conclude that multiwavelength 
high-angular-resolution observations and radiative transfer disk 
models are indispensible to understand these complex interacting 
objects and their place in the grand scheme of the (binary) evolution 
of low and intermediate mass stars.
\end{abstract}

\section{Introduction}
Based on recent surveys of the optically-bright post-AGB population 
in the Magellanic Clouds (MCs) 
\citep[][see also D. Kamath's contribution in these 
proceedings]{2011AAvanAarle,2014MNRASKamath}, the formation of disks 
around post-AGB binaries seems to be a common process. 
Indeed, in analogy with the Galactic post-AGB stars with confirmed 
disks, about 40\% of the optically bright post-AGB stars in the MCs 
have similar observational characteristics (i.e. a comparable IR excess
and photospheric depletion pattern). Binary interaction is clearly a 
key ingredient in the formation of these disks, since in the Galaxy 
such stable structures are only found around post-AGB stars in binary 
systems of typically 1~AU in separation ($P_{\rm{orb}}\sim100-3000$~d). 
The advantage of studying post-AGB stars in the MCs is that their 
distances, and hence luminosities, are well constrained, which is not the 
case for a typical Galactic source. On the other hand, to constrain the 
structure and evolution of the circumstellar environment in greater 
detail, one can better study the Galactic objects which can be spatially 
resolved with high-angular-resolution techniques. Here we focus on the 
results obtained in our recent studies of this kind. The long-term goals 
of this research are to further binary evolution theory by
\begin{itemize}
 \item  empirically constraining uncertain binary interaction processes 
related to the formation of these elusive disks,
 \item connecting the post-AGB binaries to other objects and evolutionary 
channels in the ``binary zoo,'' in search of their progenitors and progeny,
\end{itemize}
but also to study disk evolution in itself, since these objects 
\begin{itemize}
 \item form an ideal laboratory to study dust coagulation in a 
semi-stable environment,
 \item offer a unique region of parameter space to study mechanisms 
that are relevant for the formation of (circumbinary) planets.
\end{itemize}

\section{The Two Stars: the Prototypes in Hercules}
Two post-AGB systems were selected for a detailed study of the 
structure of their circum\-stellar environment, 89 Herculis 
\citep[published as][]{2013AAHillen,2014AAHillen} and AC Herculis 
(Hillen et al., in prep.). Both systems are among the brightest and 
closest post-AGB binaries and have a long history as being recognised 
as likely disk objects. \citet{1993AAWaters} postulated 89 Her to have 
a disk, based on their evidence for the binary nature of the central 
object and the observed characteristics of the circumstellar environment 
(i.e. the stability of the IR excess, the CO\,(1-0) line profile, etc.). 
Similarly, \citet{1998AAVanWinckel} concluded, based on the close 
resemblance of AC Her with the Red Rectangle (i.e., the dust mineralogy, 
CO rotational line emission, mm continuum flux, etc.), that the 
circumstellar dust and gas in this system must also be in the form of 
a circumbinary disk.  Simple radiative transfer disk models have already 
been computed for AC Her, in combination with a mineralogical study
of the mid-IR emission features, by \citep[][]{2007AAGielen}. 
Here we compare our results for the two systems.

\section{Our Tools}
\subsection{Observations: Optical Interferometry Combined with the SED}
Extensive high-quality data sets were gathered for the two objects under 
study. For both systems the spectral energy distribution (SED) was 
constructed with a wide variety of photometric data from the literature, 
combined with new photometry collected with the SPIRE instrument onboard 
the {\sl Herschel} satellite \citep[][]{SPIRESwinyard,2010AAPilbratt}, 
as well as with the archival ISO spectra.

For 89 Her, we collected multiwavelength interferometric data, with 
currently operational interferometers (the VLTI, the CHARA Array and 
the NPOI) and from the archives of the PTI and the IOTA, that cover 
the optical, near-IR and mid-IR wavelength domains. In the case of 
AC Her, only three visibility spectra were acquired with the MIDI 
instrument on the VLTI, but they are of very high quality and 
spatial resolution.

\subsection{Radiative Transfer Disk Models}
\begin{figure}[!ht]
 \caption{A diagram explaining the iterative sequence executed in the 
MCMax code which leads to the final disk structure.}
 \plotone{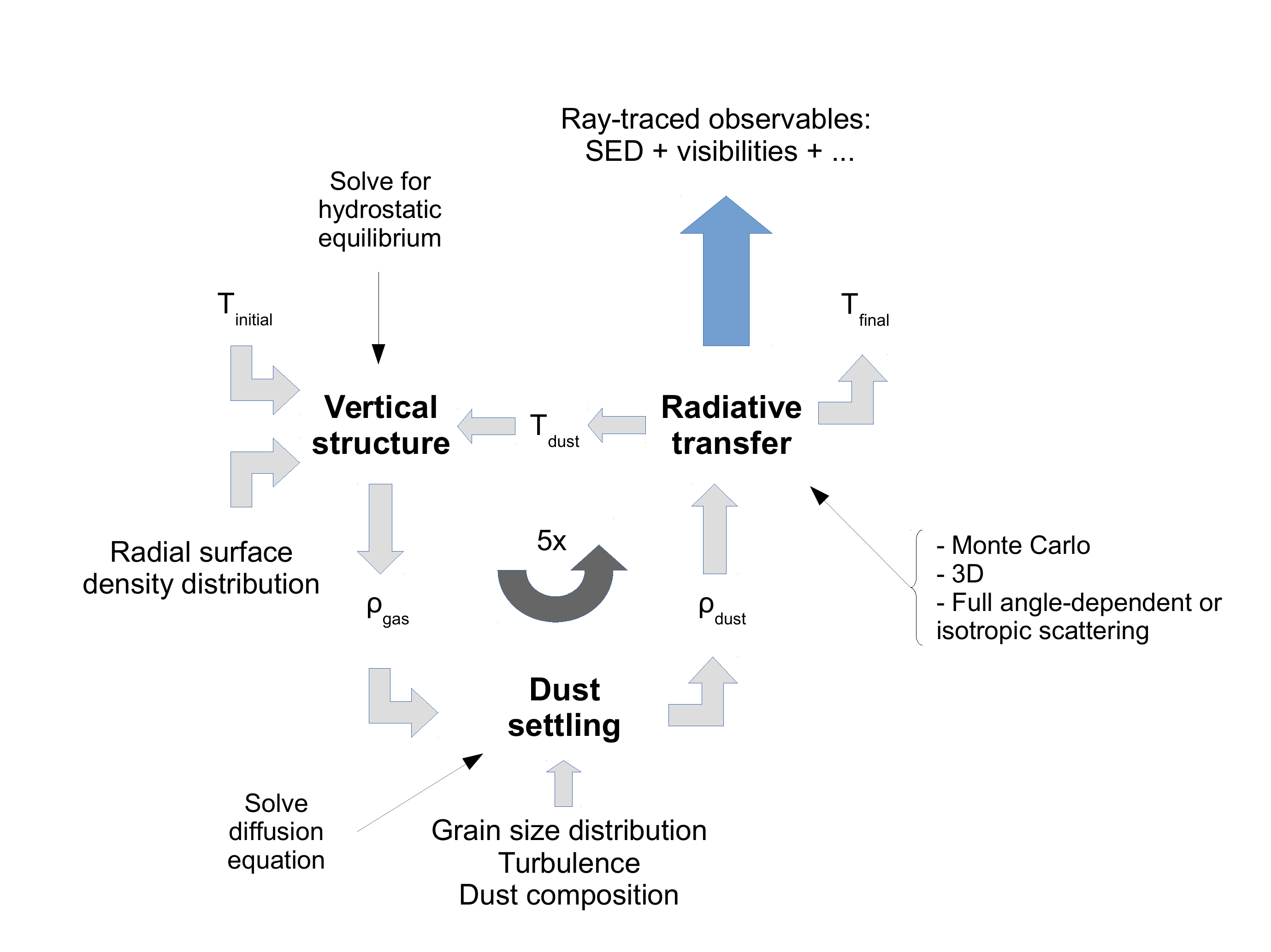}
 \label{figure:diagram}
\end{figure}

The main modeling tool used in our work is the MCMax radiative 
transfer code \citep{2009AAMin}. Being developed to model the effects 
of radiation transport through the optically thick media in 
protoplanetary disks on dust-related observables, MCMax can be very 
well applied to the circumstellar environments of evolved stars as 
well, and in particular to the disks around post-AGB binaries. 
The radiative transfer in MCMax is performed with a Monte Carlo method. 
The code, moreover, computes the vertical structure of the disk, by 
solving the equation of hydrostatic equilibrium. Finally, a grain size 
distribution can be included in the model, in combination with 
size-dependent dust settling to the disk midplane (i.e.~turbulence 
vs.~gravity, included in the form of a diffusion equation). 
An important assumption in the modeling is the way the radial surface 
density distribution is parameterized, which is typically in the form 
of a power law (with index -1 for protoplanetary disks).
Figure~\ref{figure:diagram} summarizes in the form of a diagram the 
iterative processes implemented in MCMax to arrive at a final disk 
structure and model predictions for observables 
\citep[for more details, see][and references 
therein]{2014AAHillen,2012AAMulders,2009AAMin}. 
In particular, the inclusion of mm-sized grains that are settled to 
the midplane of the disk is a great improvement with respect to 
previous radiative transfer models applied to post-AGB disks.

\section{Modeling Results: Comparing the Two Stars}
Extensive model grids were computed to explore the relevant parameter 
space. Due to the complexity of these stable, passive structures, 
there are many parameters involved in the structure computation, in 
addition to the geometric parameters like inclination and disk 
orientation on the sky (see Table~\ref{table:MCMaxmodels}). 
Not all parameters can be independently constrained based on the SED 
and interferometric data at near- to mid-IR wavelengths only. 
Therefore, we assumed certain values for specific parameters. 
These assumptions are different for the two systems, complicating 
a quantitative comparison between the resulting values for the fitted 
parameters. Nonetheless, several conclusions can be drawn with respect 
to the parameters that \textit{are} well determined and concerning the 
validity of certain assumptions/parameterizations.

\begin{table}
 \caption{Parameters of the best-fit MCMax radiative transfer disk 
models of 89~Her and AC Her. ~(F) means that the parameter was fixed 
to the given value.}         
 \label{table:MCMaxmodels}    
 \centering                
 \smallskip
 \smallskip
 \begin{tabular}{l c c}      
 \tableline
 \noalign{\smallskip}
  Parameter & Best 89 Her & Best AC Her \\   
 \tableline
 \noalign{\smallskip}
 \noalign{\smallskip}
  $M_{\mathrm{dust}}$ (M$_\odot$) & $5\,\times\,10^{-4}$ & 
$2.5\,\times\,10^{-3}$ \\
  gas/dust & 100 (F) & 1 or 10 \\
  $a_{\mathrm{min}}$ & 0.01 (F) & 0.01 (F) \\
  $a_{\mathrm{max}}$ (mm) & 10.0 (F) & 1.0 \\
  $q$ & --3.00 & --3.25 \\
  $R_{\mathrm{in}}$ (AU) & 3.75 & 34.0 \\
  $R_{\mathrm{out}}$ & 50 (F) & 200 (F) \\
  $R_{\mathrm{mid}}$/$R_{\mathrm{in}}$ & 3.0 & 2.0 \\
  $p_{\mathrm{in}}$ & --3.0 & --3.0 \\
  $p_{\mathrm{out}}$ & 1.5 & 1.0 (F) \\
  $\alpha$ & 0.01 (F) & 0.01 (F) \\    
  $i$ ($^\circ$) & 13 (F) & 50 \\
 \noalign{\smallskip}
 \tableline
 \noalign{\smallskip}
 \end{tabular}
\end{table}

First, it is striking that both systems require a surface density 
parameterization of two joined power laws to explain the data. 
The interferometric data (in the near-IR for 89 Her and in the mid-IR 
for AC Her) require a smoother intensity distribution than can be 
provided with a single power-law model 
\citep[see][for a detailed discussion]{2014AAHillen}. 
Second, it is apparent that in both systems the grain size distribution 
power-law index is larger than --3.5, the value often assumed for 
protoplanetary disks. For post-AGB disks the inclusion of large grains 
is thus clearly very important.  Third, our derived dust masses are 
rather high. In the case of 89 Her, our dust mass is a factor of five 
larger than the value estimated from the measured gas mass (from CO 
rotational lines) by \citet{2013AABujarrabalB} combined with a standard 
gas/dust ratio of 100. This we judge to be within the errors of both 
methods, especially given the different assumed distance.  In the case 
of AC Her, the gas/dust ratio is also a fit parameter in our models 
because it affects the settling of dust particles and thus the shape 
of the inner rim, and hence the interferometric data. 
Our dust mass for this system is a factor $\sim$3 larger than the 
total gas mass that was found by \citet{2013AABujarrabalB}. With the 
indication for a gas/dust ratio smaller than 100 (the best-fit value 
is $\sim$1-10) from our models and our larger distance, this discrepancy 
is within what the respective errors allow.  Nevertheless, it would be 
interesting to check whether the CO lines are affected by optical depth 
effects, to see whether our modeling might be biased by any of our 
assumptions or simplifications.
Only by combining the various data sets can this be resolved.
Finally, the foremost distinction between the two systems are their 
vastly different inner radii. The hottest dust in 89 Her 
coincides rather well with the expected dust condensation radius. 
In AC Her the inner rim is located much further out, almost an order 
of magnitude beyond the dust condensation radius, categorizing it as 
a ``post-AGB transitional disk.'' The origin of this large inner hole 
is yet unexplained.  It is interesting that AC Her combines a rather 
large disk mass with a seemingly evolved inner disk. On the other hand, 
89 Her has a relatively massive large-scale outflow, well-resolved in 
CO rotational lines, despite its inner radius coinciding with the dust 
condensation radius. Such an outflow is not yet detected in AC Her. 
Are different disk dispersal mechanisms responsible for the current 
state of the two systems?

\section{Conclusions}
Optical interferometry is a powerful technique to trace the inner 
regions of dusty disks. Radiative transfer modeling techniques of 
optically thick media have come of age in the past decade and can now 
be successfully applied to the circumbinary disks around post-AGB 
binaries.  Combining these tools allows us to constrain the elusive inner 
disk regions in great detail. We have shown for two systems that a large 
set of state-of-the-art observations can be well matched with these 
models, but that the resulting parameter values raise several questions 
concerning their evolution.

The works presented here, and even published throughout the literature, 
have only scratched the surface of what is feasible. With the 
2nd-generation instruments coming online on the VLTI in the coming years, 
and ALMA almost fully operational, more exciting results can be expected 
in the coming decade. By tracing the complex structures and matter 
streams in a large number of post-AGB objects, we hope to connect these 
peculiar systems to specific populations of stars from which they 
originate and into which they evolve. An exciting time lies ahead!  

\bibliography{Hillen1}

\question{Posch} You mentioned amorphous carbon and iron as continuum 
opacity sources in your models. How well is the abundance of amorphous 
carbon or iron constrained by your models? What would happen if you'd 
set the abundances of Fe and amorphous carbon to zero in your models?

\answer{Hillen} The abundance of amorphous carbon or metallic iron 
cannot be constrained with the data and models that we have. Other 
parameters, like the grain size distribution power law index, can 
mimic effects of varying opacities/abundances.

\end{document}